\documentclass{article}
\usepackage[utf8]{inputenc}
\usepackage[T1]{fontenc}
\usepackage{amsmath}
\usepackage{comment}
\usepackage{graphicx}
\usepackage{xcolor}
\usepackage{geometry}
\usepackage{authblk}
\newgeometry{vmargin={1in}, hmargin={1in}}
\usepackage{lineno}

\usepackage{hyperref}

\title{Spatial Confinement Affects the Heterogeneity and Interactions between Shoaling Fish}

\author[a]{Gabriel Kuntz}
\author[b]{Junxiang Huang}
\author[a]{Mitchell Rask}
\author[a]{Alex Lindgren-Ruby}
\author[a]{Jacob Y. Shinsato}
\author[b]{Dapeng Bi}
\author[a,c,*]{A. Pasha Tabatabai}

\affil[a]{Seattle University, Department of Physics, Seattle WA, 98122, USA}
\affil[b]{Northeastern University, Department of Physics, Boston, MA, 02115, USA}
\affil[c]{California Polytechnic State University, Physics Department, San Luis Obispo CA, 93410, USA}

\affil[*]{pashatab@calpoly.edu}

\begin{document}
    
\flushbottom
\maketitle

\thispagestyle{empty}

\begin{abstract}
Living objects are able to consume chemical energy and process information independently from others. However, living objects can coordinate to form ordered groups such as schools of fish. This work considers these complex groups as living materials and presents imaging-based experiments of laboratory schools of fish to understand how this non-equilibrium activity affects the mechanical properties of a group. We use spatial confinement to control the motion and structure of fish within quasi-2D shoals of fish. Using image analysis techniques, we make quantitative observations of the structures, their spatial heterogeneity, and their temporal fluctuations.
Furthermore, we utilize Monte Carlo simulations to replicate the experimentally observed area distribution patterns which provide insight into the effective interactions between fish and confirm the presence of a confinement-based behavioral preference transition. In addition, unlike in short-range interacting systems, here structural heterogeneity and dynamic activities are positively correlated as a result of complex interplay between spatial arrangement and behavioral dynamics in fish collectives.
\end{abstract}

\section*{Introduction}

Nature provides fantastic examples of complex collective behaviors on many length scales in order to accomplish certain tasks. For example, cells within tissues coordinate to successfully close wounds~\cite{ajeti_wound_2019,brugues_forces_2014}, ants build structures to overcome obstacles~\cite{ko_small_2022, graham_optimal_2017}, and fish form cohesive groups to improve computations about their environment~\cite{berdahl_emergent_2013,hein_evolution_2015}. In each of these examples, the interactions between individuals lead to function on a larger scale.

Understanding the details of the interactions between individuals within these complex groups is an active area of research~\cite{katz_inferring_2011, sridhar_geometry_2021}. Previously, it has been shown that metric interactions, where constituents within a certain distance interact, qualitatively capture the collective behaviors seen in flocking~\cite{vicsek_novel_1995}. However, closer inspections in a variety of species suggest that the true interactions are more likely visual, topological~\cite{ballerini_interaction_2008, camperi_spatially_2012, kumar_flocking_2021}, or more complicated~\cite{torney_inferring_2018}. The pursuit of understanding these interactions is valuable to understanding fundamental problems in complex systems.

However, this connection between constituent interactions and bulk behavior parallels the language used to describe and design materials. Examples of this include tuning the interaction strength between colloids to influence colloidal gel rheology~\cite{shah_viscoelasticity_2003,lu_gelation_2008, hsiao_role_2012, nabizadeh_life_2021}, the interaction specificity within DNA hybridized colloids and material structure~\cite{rogers_direct_2011, wang_crystallization_2015, hayakawa_geometrically_2022}, and the relative physical parameters within models of epithelial tissue and tissue fluidity~\cite{bi_density-independent_2015,yang_correlating_2017}. Therefore, this search for the relationship between interactions and bulk behavior is also critical for defining collective systems as living materials in-and-of themselves. 
In particular, these living materials are fundamentally non-equilibrium due to the local consumption of energy by each entity, and definitions of the mechanical~\cite{lee_quantifying_2013, ni_intrinsic_2015, van_der_vaart_mechanical_2019} or thermodynamic~\cite{takatori_swim_2014, bowick_symmetry_2021} properties of these systems will help determine the material possibilities of these types of systems.

In this paper, we aim to understand the mechanical properties of a quasi-2D living material - groups of aquarium fish within the lab; for simplicity, we confine fish to thin volumes of water. We make quantitative observations of groups of swimming fish using image analysis techniques to identify fish positions and trajectories. We track fluctuations of structures at the local and group level as self-generated deformations.
We find that we can control the motile behaviors of these fish by varying the level of spatial confinement and that this change in individual motion is correlated with a changing heterogeneity of the group. 
In addition, we employ Monte Carlo simulations to recreate the complex dynamics of fish interactions and examine the effects of varying spatial constraints on the group's mechanical properties. By using simulations to replicate experimentally measured distributions, we infer effective interactions between individuals. Through a wide variety of metrics, we observe a behavioral transition as a function of spatial confinement which correlates with changes in structural heterogeneity. 
In contrast to colloidal gels involving short-range interactions, the simulations uncover a positive correlation between structural heterogeneity and dynamic activities. 

\section*{Results}
\begin{figure}[]
 \centering
\includegraphics[width=.3\linewidth]{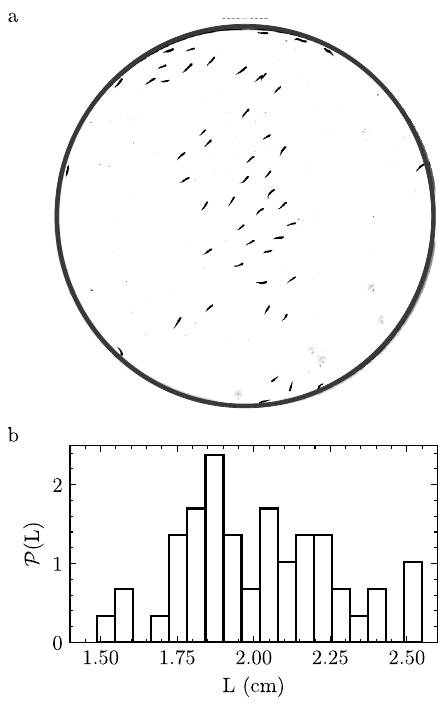}
 \caption{(a) Image of 50 fish within arena with radius $R = 44.25$cm. Arena boundary outlined for clarity. (b) Probability distribution of fish mid-line length ($L$) for 50 fish in (a).}
 \label{fig:L_Hist}
 \end{figure}

Cardinal tetra fish are imaged moving freely within quasi-2D cylindrical arenas with a depth of 1.5cm$\pm$0.1cm (Methods). Fish positions $\Vec{r}$ are identified using the open-source software Trex~\cite{walter_trex_2021} which identifies fish body cross-sections through image contrast. We also use this software to connect fish positions in time to build trajectories $\Vec{r}(t)$ and calculate instantaneous velocities $\Vec{v}(t)$ for each fish (Methods). Since we now have the projected area of each fish, we use the mid-line length to characterize fish size. We find fish have a length of $L=2.0 \pm 0.2$ (mean $\pm$ stdev) (Figure ~\ref{fig:L_Hist}).

\begin{figure}[]
\centering
\includegraphics[width=\linewidth]{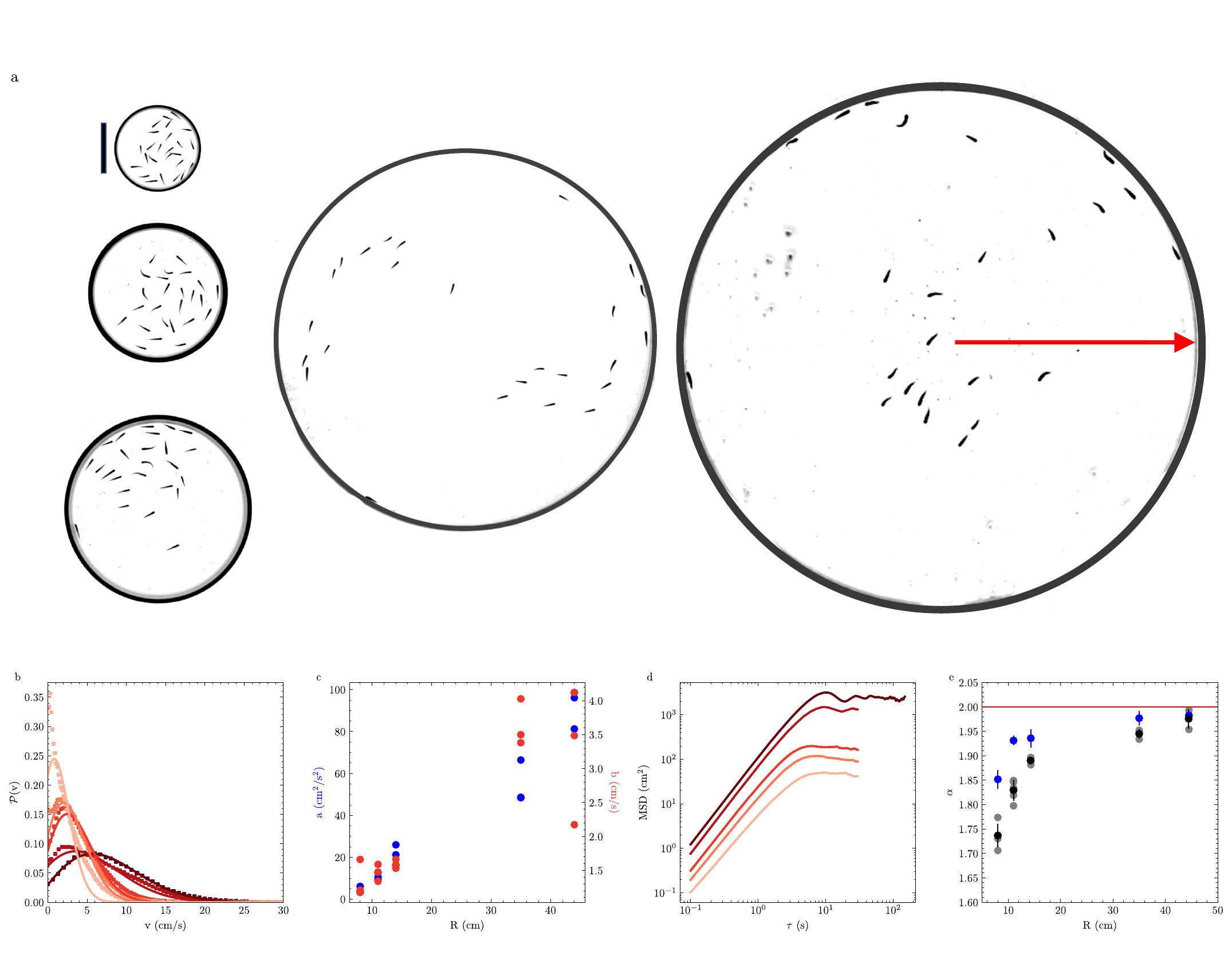}
 \caption{\textbf{Confinement affects motion.} (a) Images of arenas with radii $R = 8$cm, 11cm, 14cm, 34.25cm \& 44.25cm containing 25 fish. Arena boundaries were added for clarity. The red arrow denotes radius $R$. The scale bar is 10cm.
 (b) Example probability distributions of fish speed ($v$) for 25 fish in the differently sized arenas (markers) and best fit Gaussian functions (lines).
 (c) Fitting parameters $a$ (left-blue) and $b$ (right-red) for the lines in (b).
 (d) Mean squared displacement (MSD) as a function of elapsed time ($\tau$) for fish in five different arenas. 
 (e) The short-time power-law slope of MSD ($\alpha$) as a function of radii ($R$). Experimental duplicates of $\alpha$ are plotted in grey ($N=3$ for each), with the mean plotted in black. $\alpha$ for the shoal center of mass motion (blue). Error bars are one standard deviation. The plotted color darkens as $R$ increases in (b) and (d). 
 }
 
 \label{fig:msd_Fig}
 \end{figure}

We record videos of 25 fish within arenas of different radii ($R$) to investigate the effects of confinement (Figure~\ref{fig:msd_Fig}a). Altering the radius $R$ adjusts the global area density $\rho_g = N/(\pi R^2)$. However, it is crucial to note that fish can exhibit significant local density fluctuations since they do not uniformly occupy the entire space~\cite{makris_fish_2006,Becco_physica_A_2006}. 
We observe that arena size influences the probability distribution of speeds ($v$), where decreasing $R$ biases the distributions towards slower speeds (Figure~\ref{fig:msd_Fig}b). This broad distribution of speeds is consistent with the stop-start motion associated with fish motility~\cite{li_intermittent_2023}. We fit each speed probability distribution to a modified Rayleigh distribution. 
\begin{equation}\label{eq:Gaussian}
    P(v) = \frac{v+b}{a} e^{-(v+b)^2/(2a)}
\end{equation}
where $a$ and $b$ are fitting parameters associated with the width and shift of the distribution accordingly. We find that both of these parameters increase with $R$ (Figure\ ~\ref{fig:msd_Fig}c). This functional form was chosen to resemble the 2D Maxwell-Boltzmann distribution in an attempt to make parallels between the motion of molecules at thermodynamic equilibrium and the motion of fish out-of-equilibrium; the connection is elaborated on further in the Discussion Section.

While Figure~\ref{eq:msd}b indicates that the magnitude of motion is affected by $R$, it does not describe the persistence of motion. As such, we calculate the mean squared displacement (MSD) 
\begin{equation}\label{eq:msd}
 \mathrm{MSD} (\tau) =\left< (\vec{r}(t + \tau) - \vec{r}(t))^2 \right> 
 \end{equation}
by comparing positions $\vec{r}(t)$ as a function of elapsed time ($\tau$) for each fish; we average the MSD from all fish within an experiment to generate a single ensemble-averaged MSD (Figure~\ref{fig:msd_Fig}d, Methods).
For large $\tau$, the MSD turnover and plateau are set by the finite size of the arena.
For small $\tau$, we observe power-law scaling (i.e. $\mathrm{MSD} \sim \tau ^{\alpha}$) where $\alpha$ characterizes the type of motion. We find that $\alpha$ depends on confinement, demonstrating that fish motion is super-diffusive ($1 < \alpha < 2$) and approaches ballistic motion ($\alpha\xrightarrow{}2$) as containers get larger (Figure~\ref{fig:msd_Fig}e). The MSD of the shoal's center of mass has similar arena-size dependent values of $\alpha$ (Figure~\ref{fig:msd_Fig}e, Supplemental Figure 1).

\begin{figure}[]
 \centering
 \includegraphics[width =\linewidth]{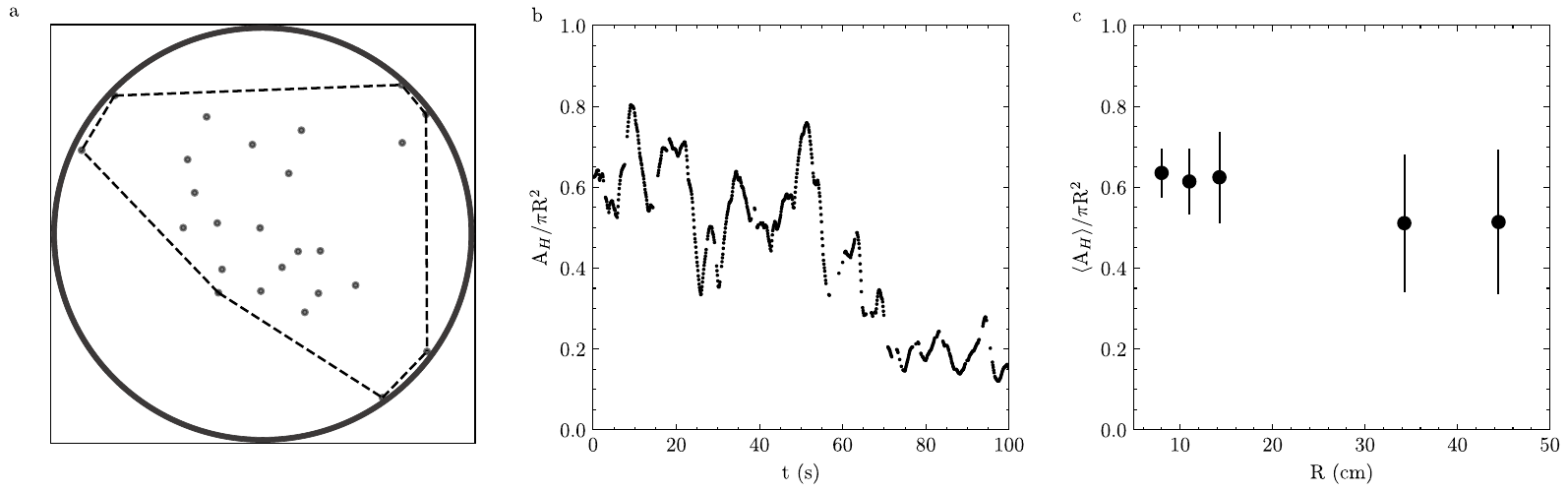}
 \caption{\textbf{Shoal area fluctuates in time.} 
 (a) Convex hull of 25 fish in $R=44.25$cm arena. Fish positions are grey markers. Convex hull is defined by the black dashed line. The image width is $2R$ with the arena boundary outlined for clarity. (b) Example of convex hull area ($A_H$) normalized to arena area ($\pi R^2$) as a function of time for 25 fish in R=34.25cm arena.
 (c) Time average convex hull area $\left<A_H\right>$ normalized by arena area for arenas of different radii. Error bars are one standard deviation.} 
 \label{fig:Hull}
 \end{figure}

We next asked how these differences in motion affected the organization of fish in groups. To investigate the effects of confinement on structural and material properties of the shoal we use fish positions $\vec{r}(t)$ to calculate the time-varying convex hull (Figure~\ref{fig:Hull}a, Methods) which we use to define the overall geometric size of the group. The area of the convex hull $A_{H}$ fluctuates considerably over time, signifying that the group is exploring different fractions of the space (Figure~\ref{fig:Hull}b). While $A_{H}$ fluctuates in time, the time-averaged fraction of space occupied by the shoal $\left<A_{H}\right>/\pi R^2$ is consistent across different confinements (Figure~\ref{fig:Hull}c); on average, the shoal will fill the space to an equal extent regardless of the amount of space it has available to it (Supplemental Figure 2).

\begin{figure}[]
 \centering
 \includegraphics[width =\linewidth]{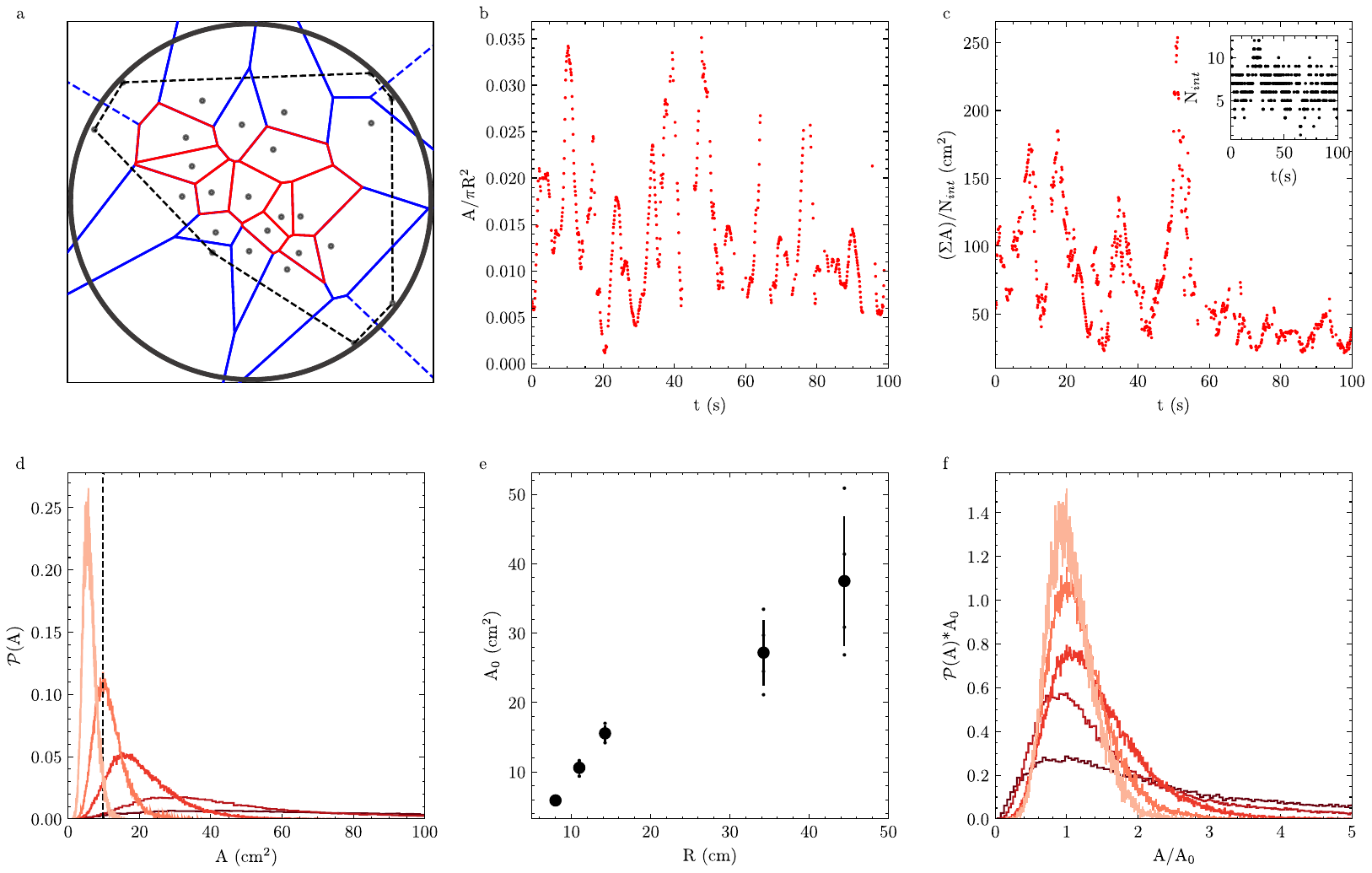}
 \caption{\textbf{Confinement affects local fish packing.} (a) Voronoi tessellation (polygons) of 25 fish (grey markers) in $R=44.25$cm arena. Internal fish (red polygons) are a subset of all fish and have all vertices within the convex hull (black dashed line). The image width is $2R$ with the arena boundary outlined for clarity. (b) Example time evolution of a single fish Voronoi area ($A$) normalized by arena area ($\pi R^2$) in $R=34.25$cm arena.
 (c) Sum of internal areas ($\Sigma A$) normalized by number of internal fish ($N_{int}$) as a function of time for data in (b). (c-inset) Number of internal fish ($N_{int}$) as a function of time for data in (b). (d) Probability distributions ($P(A)$) of internal areas ($A$) for 25 fish for different arena sizes. The vertical dashed line indicates the peak of a distribution $A_0$. (e) $A_0$ as a function of arena radii ($R$). Bold circles are averages across different fish groups, small data points are individual experiments and error bars are one standard deviation. (f) Scaled probability distributions from (d). Colors in (d) and (f) darken with increasing $R$.}
 \label{fig:Voronoi}
 \end{figure}

The size of a shoal characterized by $A_H$ is prone to bias by fish that do not move with the group. Therefore, we define a local measurement of the space occupied by individual fish by calculating the Voronoi tessellation using the fish positions $\vec{r}$. The Voronoi cell for a particular fish is the space that is most proximal to a fish; this acts as an amorphous unit cell. We restrict our structural analysis at each frame to fish that have Voronoi cells completely enclosed within the convex hull (Figure~\ref{fig:Voronoi}a); these `internal' fish have Voronoi areas which are both closed and do not drastically change with small neighbor movements. We calculate the areas of each of these Voronoi cells $A$ for individual fish and find that they fluctuate through time as well. In  Figure~\ref{fig:Voronoi}b, we show an example Voronoi area that fluctuates by an order of magnitude in area over the plotted observation window. We also note that the data point frequency is not constant over the observation window; no Voronoi area is calculated for this fish if it fails to be an `internal' fish or if we cannot uniquely identify all fish in a particular frame.

The number of these internal fish $N_{int}$ varies widely over time (Figure~\ref{fig:Voronoi}c inset) and each internal fish has an average of approximately six topological neighbors, defined as neighbors that share a Voronoi edge (Supplemental Figure 3).
Therefore, to further understand the relationship between local fluctuations in $A$ and shoal level fluctuations, we estimate the time-varying net size of all internal fish $\sum A / N_{int}$ (Figure~\ref{fig:Voronoi}c). We note that the time courses of the $A_H$ (Figure~\ref{fig:Hull}b), $A$ (Figure ~\ref{fig:Voronoi}b), and $\sum A / N_{int}$ (Figure~\ref{fig:Voronoi}c) all come from the same experiment and are plotted over the same range of time. The relative size of fluctuations within $A_H$ and $\sum A / N_{int}$ are qualitatively similar. However, these are not equivalent to the fluctuations of $A$ over the same time period. Therefore, the group and the individual area fluctuations are not mirrored, and fish areas do not all uniformly expand or shrink collectively.

Since shoals occupy larger spaces in larger arenas (Figure~\ref{fig:Voronoi}c), the Voronoi region associated with each fish must also vary with $R$. Indeed, we see this dependence in the probability distributions of $A$ for all internal fish within an experiment, where increasing $R$ biases the distribution to larger areas $A$ (Figure~\ref{fig:Voronoi}d). We define the modal area $A_0$ which increases with arena size, however, it does not increase $\sim R^2$ as would be expected for a 2D gas (Figure~\ref{fig:Voronoi}e).

By comparing the normalized probability distributions ($P(A)*A_0$ vs $A/A_0$), we show that the fluctuations observed are statistically similar between small arena sizes yet vary significantly for larger arenas (Figure~\ref{fig:Voronoi}f); this is consistent with the arena size dependence of MSD scaling in Figure~\ref{fig:msd_Fig}. We also show that these distributions are not equivalent to distributions made from randomly generated points, consistent with the fact that fish are not randomly occupying space (Supplemental Figure 4).

\begin{figure}[h]
 \centering
 \includegraphics{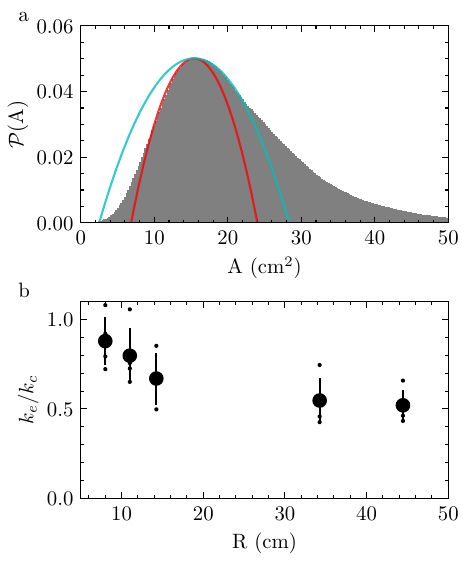}
 \caption{\textbf{Confinement affects area distribution asymmetry.}(a) Example probability distribution of internal fish voronoi areas ($A$) for $R=14$cm arena (143,963 instances). Parabolic fits to $A<A_0$ (red) and $A>A_0$ (cyan) associated with compression and expansion, respectively. (b) Ratio of coefficients of expansion $k_e$ and compression $k_c$ found from fit in (a) as a function of arena radius ($R$). Small markers are individual experiments, and the large black markers are mean with one standard deviation error bar.} 
 \label{fig:Comp&Exp}
 \end{figure}

Upon inspection, the probability distribution of observed internal Voronoi areas is not symmetric about the mode $A_0$ (Figure~\ref{fig:Comp&Exp}a). Here, we take an approach that is similar to the computational modeling of cells in tissues via the Vertex and Self-Propelled Voronoi models where deviations of a cell from a modal area are associated with an energy cost for that cell~\cite{bi_density-independent_2015,yang_correlating_2017}. To understand the underlying dynamics that result in the asymmetric distributions in Figure~\ref{fig:Voronoi}d, we fit two separate parabolic functions to each distribution such that
\begin{equation}
P(A_0) - P(A)\sim 
\begin{cases}
    k_c (A-A_0)^2, & \text{if } A< A_0\\
    k_e (A-A_0)^2, & \text{if } A> A_0
\end{cases}
\label{eq:kce_define}
\end{equation}
with constants $k_c$ and $k_e$ associated with compression and expansion, respectively, of Voronoi areas away from the modal area $A_0$ (Figure~\ref{fig:Comp&Exp}a, Methods). The ratio $k_e/k_c$ of these constants is a signature of the effective interactions between fish, which varies with $R$ (Figure~\ref{fig:Comp&Exp}b). 


\section*{Simulations}

\begin{figure}[h]
\centering
\includegraphics[width=\textwidth]{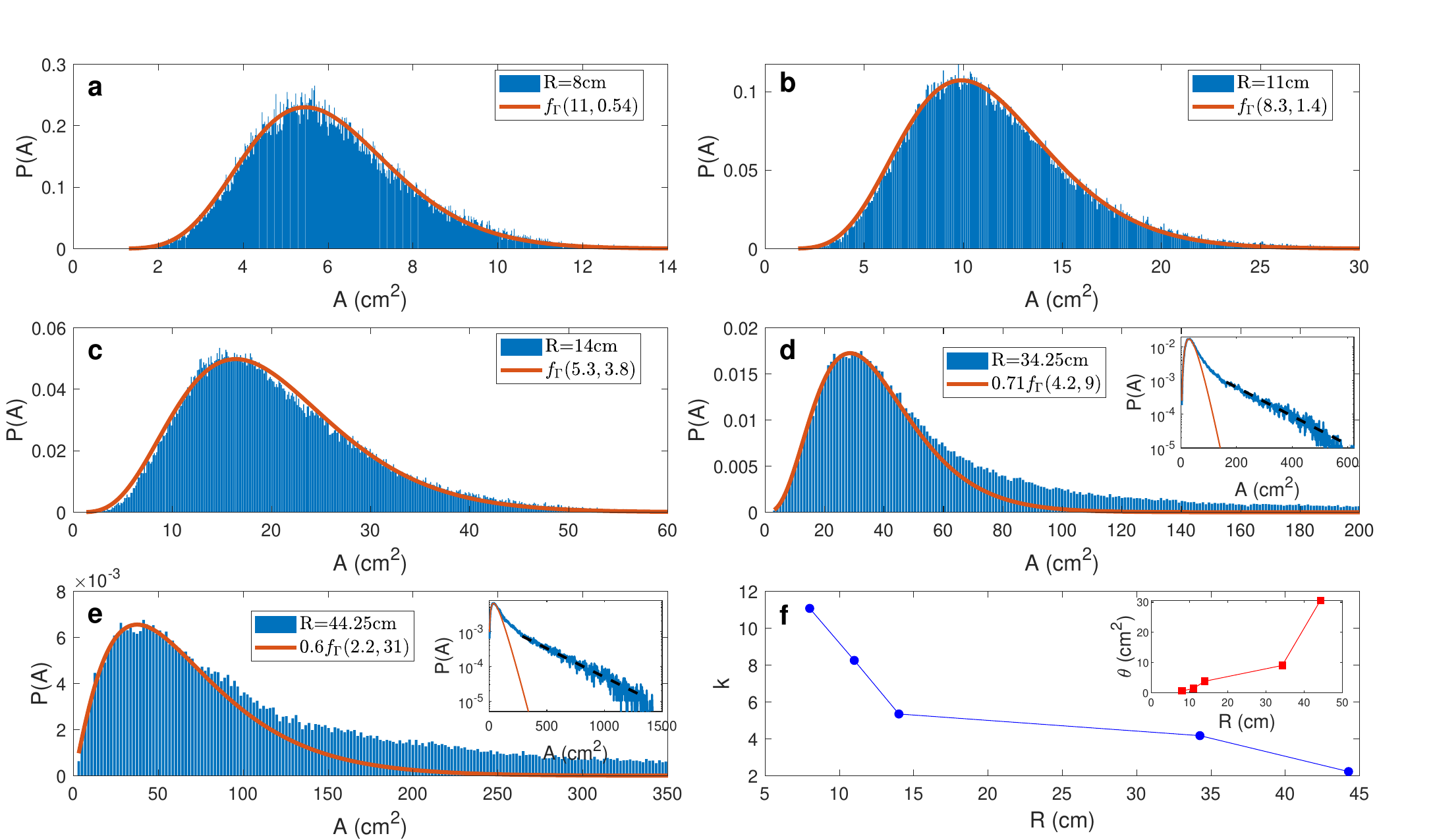}
\caption{\textbf{The heterogeneity of internal Voronoi areas increases as the density is decreased.} (a-e) Experimental area $A$ distributions are fit to Gamma distribution (red) for $R=8, 11, 14,34.25 \mathrm{and\ } 44.25$\ cm radii arenas.
$f_{\Gamma}(k,\theta)$ denotes the PDF of Gamma distribution defined in Equation \eqref{eq:gamma_pdf}. (d,e) At $R=34.25\mathrm{cm}$ and $44.25\mathrm{cm}$ the Gamma distribution fit is done for $A<1.5A_0$ and a normalization factor is applied to $f_{\Gamma}(k,\theta)$ to align the modes. Insets in (d,e) show exponential tails (black dashed lines) with decay parameters of $0.0098$ and $0.0038\mathrm{cm}^{-2}$, respectively. (f) The shape parameter $k$ as a function of arena radii $R$. (f-inset) The scale parameter $\theta$ as a function of arena radii $R$.
}
\label{fig:gamma_fit}
\end{figure}

Previous literature has established that the area distribution of cells within two-dimensional random Voronoi networks adheres to a Gamma distribution~\cite{kiang1966random}. The probability density function (PDF) for such a distribution is mathematically represented as:
\begin{equation}
f_{\Gamma}(x) = \frac{x^{k-1}e^{-x/\theta}}{\theta^k \Gamma(k)},
\label{eq:gamma_pdf}
\end{equation}
where $k$ and $\theta$ are the shape and scale parameters, respectively, while $\Gamma(*)$ denotes the gamma function. For random Voronoi networks, a shape parameter $k=3.63$ is reported to yield the best fit to the observed cell area distribution~\cite{weaire1986distribution}. In addition, for hard disks, the distribution of Voronoi free area, which is the difference between the actual Voronoi cell area and the minimum cell area at close packing, is well described by a Gamma distribution with $k$ between $3$ to $4$~\cite{kumar_flocking_2014}.

Intriguingly, despite the non-equilibrium and non-random nature of the fish collective, the internal area distributions at smaller radii $R$ conform closely to a Gamma distribution (Figure~\ref{fig:gamma_fit}a-c). The shape parameter $k$, which governs the distribution asymmetry and tail behavior, is greater than the non-interacting limit ($k=3.6$).
Large $k$ values represent a more symmetric bell-shaped curve indicative of low heterogeneity and a tendency for cell areas to aggregate around the mean. This pattern suggests that interactions between fish within small arenas lead to a more homogeneous structural arrangement; this reduces variability and allows each fish to navigate and occupy space more effectively. As the arena radius $R$ increases, there is a clear increase in the Voronoi area heterogeneity with two distinct signatures: a decrease in the $k$ value and the emergence of an exponential tail in the distribution (Figure~\ref{fig:gamma_fit}d-f). 
At large arenas such as $R=34.25$cm or $44.25$cm, the internal areas initially adhere to a Gamma distribution up to a threshold around $1.5A_0$. Beyond this point, the distribution exhibits strong exponential tails indicative of highly heterogeneous shoal structure and significant probabilities of finding large cells and large local density fluctuations. This phenomenon bears resemblance to the behavior observed in granular aggregates with capillary interactions, while the fish aggregate is unique in its ability to adapt the $k$ value, which is obtained by fitting the bulk of the distribution to a Gamma distribution and is observed to change across a broad range~\cite{berhanu2010heterogeneous}.

The structural heterogeneity and dynamics are frequently interlinked. In colloidal gels formed through short-range attractive interactions, an increase in interaction strength leads to an increase in structural heterogeneity and dynamical arrest~\cite{dibble2006structure, hsiao2015metastable}. However, our observations in fish collectives present a contrasting scenario. Due to the long-range nature of their interactions, both structural heterogeneity and dynamical activities escalate with a greater radius $R$. In these expansive arenas, local densities experience more pronounced fluctuations which result in varied behavioral patterns: some fish form tightly-knit compact shoals while others simultaneously and independently navigate the container. 

\begin{figure}[h]
\centering
\includegraphics[width=\textwidth]{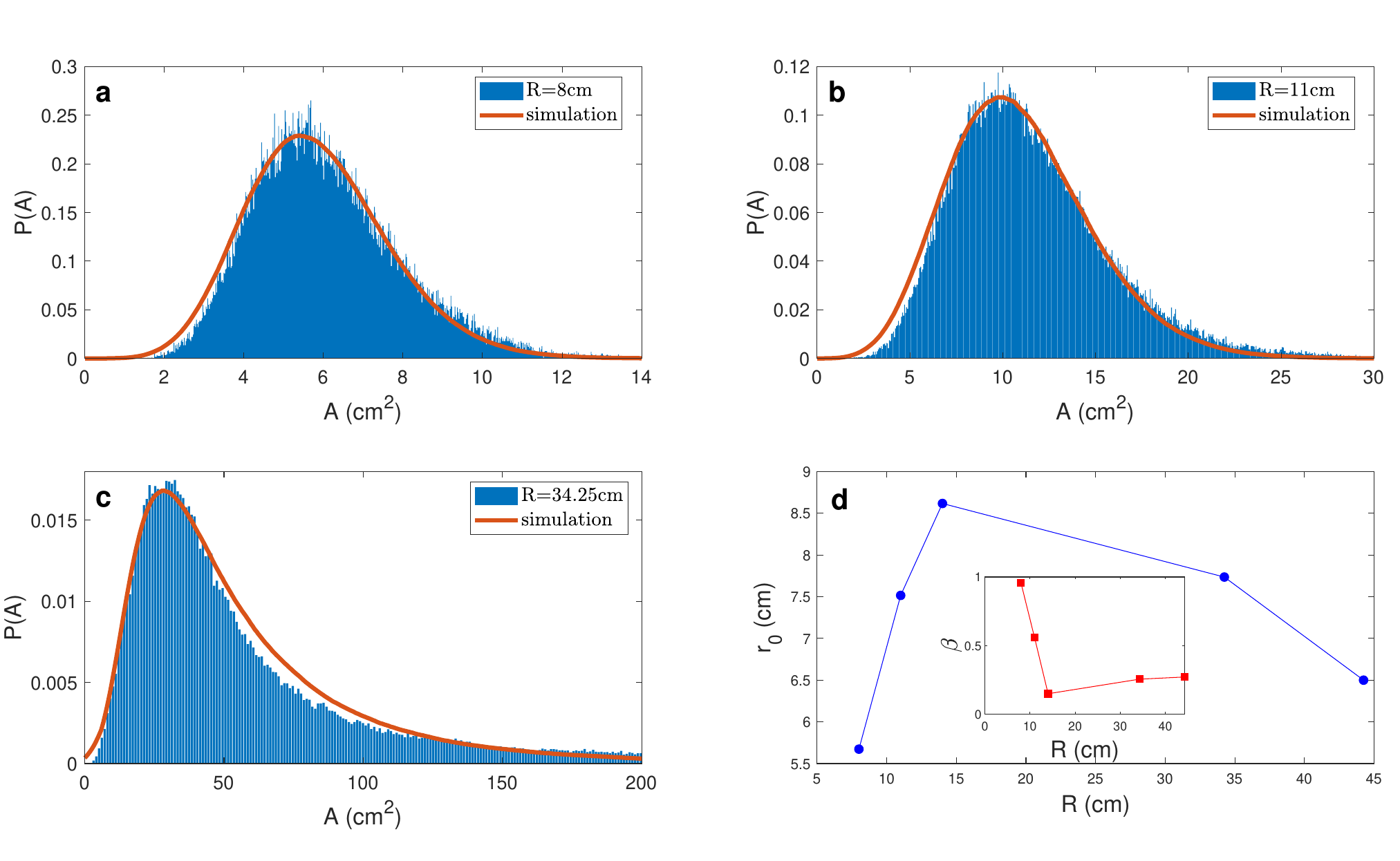}
\caption{\textbf{Interaction strength and length scale are influenced by confinement.} 
(a-c) Sample comparisons between experimental internal area distributions $A$ and Monte Carlo simulation (red line) for radii $R=8$cm, 11cm, and 34.25cm. (d) The preferred distance $r_0$ and (d-inset) parameter $\beta$ as a function of arena radii $R$ from Monte Carlo simulations.}
\label{fig:mc_fit}
\end{figure}

To elucidate the influence of the arena radius on fish interactions and the interplay between heterogeneity and dynamical activities, we implement a Monte Carlo model that simulates the fish positions with different arena radii and $\beta=1/k_B T$. At each simulation step $n$, we randomly select a fish, indexed as $i$, and propose its subsequent position as $X_{i, n+1} = X_{i, n} + \sigma \delta X_{i, n}$, where $X_{i, n}$ represents the position of fish $i$ at step $n$, $\sigma$ is the step size, and $\delta X_{i, n}$ is Gaussian white noise with unit variance. The Morse potential is adapted to emulate the complex dynamics of fish schooling by depicting the attractive and repulsive inter-fish forces. The potential is expressed as
\begin{equation}
U(r) = \left( 1 - e^{1 - r/r_0} \right)^2,
\label{eq:morse}
\end{equation}
where $r$ is the actual distance between two fish, and $r_0$ is the preferred distance. A proposed move that results in an energy change $\Delta E$ is accepted with probability $min(1, e^{-\beta \Delta E})$. We conducted the simulation with a step size $\sigma=0.4$ over $10^7$ iterations, recording the internal area metrics every $100$ step. As $\beta$ approaches zero, the influence of interactions among fish vanishes, corresponding to the ideal gas limit. Within this limit, the statistical properties of fish under different confinements are essentially the same, except for a scaling factor determined by the arena's radius. For example, the internal Voronoi areas adhere to a Gamma distribution, characterized by a constant shape parameter $k\sim 3.75\pm 0.1$ and a scale parameter $\theta(R)=0.029~R^2$, which leads to a quadratic relationship between $\left<A\right>$ or $A_0$ and $R$. To find out the parameters that best describe the experiments, we systematically sweep the parameter space of $r_0$ ranging from $0.1$ to $20$ and $\beta$ ranging from $0$ to $2$ for different simulations, and the resulting area distributions are then statistically analyzed using the chi-squared method to ascertain the optimal values for $r_0$ and $\beta$ corresponding to each radius $R$. As shown in Figure \ref{fig:mc_fit}, despite its simplified nature, the Monte Carlo simulations successfully reproduce the experimentally observed area distributions. This indicates that the core principles and rules embedded in the Monte Carlo model are effective in capturing the essential dynamics of fish interactions.

Interestingly, the preferred distance $r_0$ and $\beta$ display a non-monotonic variation as a function of $R$, with a transition point located between $R=14$cm and $34.25$cm. This non-monotonic variation suggests a complex interplay between individual space requirements and the benefits of social interactions. In small arenas, the high density compels fish to maintain a small $r_0$. When more space is available, fish tend to increase their preferred distance $r_0$ to avoid overcrowding and reduce stress, with a decreasing $\beta$ indicative of increased activities. With sufficient space, however, behavior changes and a decrease in $r_0$ imply a shift toward preserving the advantages of schooling such as enhanced communication and collective vigilance. Such a transition explains the dramatic decrease of $A_0/R^2$ for the big radii in Figure~\ref{fig:Voronoi}e, in contrast to a constant ratio in the non-interacting limit or $\beta=0$. The presence of a transition point indicates a threshold at which the fish alter their spacing behavior, possibly to balance the conflicting needs for individual space and group affiliation as a strategic response to maximize the evolutionary benefits of schooling.

\section*{Discussion}
In this manuscript, we have treated a quasi-2D shoal of fish as a living material, and we have demonstrated the ability to control the average motion of individuals as well as the structures present within the group simply by changing the extent of confinement while keeping the number of individuals within the group constant.

In describing the distribution of speeds in Figure~\ref{fig:msd_Fig}, we found that a Rayleigh function (Equation~\ref{eq:Gaussian}) described our data well. This is reminiscent of a 2D Maxwell-Boltzmann distribution and suggests that this experimental system can teach us something about energy usage within non-equilibrium systems. To make this analogy, we consider the traditional Maxwell-Boltzmann distribution which describes the speed $v$ of molecules with mass $m$ of a gas at a given temperature $T$. In this form, $k_BT$ is the thermal energy scale where $k_B$ is the Boltzmann constant. If we consider $a=k_BT/m$ and that $b$ is a fitting parameter for offset speed in Equation~\ref{eq:Gaussian}, then $a$ is analogous to the amount of energy inputted. Traditionally this energy would be via thermal fluctuations per particle, however, $a$ is not thermal in origin. Instead, this energy input comes from the energy usage of the fish towards swimming and is therefore a measure of non-equilibrium activity.
This trend of changing effective energy with confinement is also observed through the interaction energy within Monte Carlo simulations, where $\beta(R)$ in Figure~\ref{fig:mc_fit}d is similar to $1/a$ in Figure~\ref{fig:msd_Fig}c.

Extending this non-equilibrium thermodynamic analogy further, the increase in non-equilibrium activity $a$ with system size $R$ is correlated with a decrease in global density $\rho_g$. This interpretation of Figure~\ref{fig:msd_Fig}b,c suggests that fish consume more energy while swimming in the larger arenas. Therefore, our observed trend is similar to an ideal gas at constant pressure: gas molecules must have more thermal energy to maintain a constant pressure if there are fewer molecules. 

Drawing on statistical mechanics, we make the analogy that the dynamics of a material $\left<v^2\right> \sim 1/\beta$. Our simulations indeed show that an increase in $\beta$ which represents a decrease in motion, is concurrent with decreases in structural heterogeneity (larger shape parameter $k$). Therefore, we again suggest that an increase in structural heterogeneity is correlated with an increase in dynamics within fish shoals.

We observe a decoupling of the local fluctuations in fish areas with the group level fluctuations in size, indicating that expansions and contractions of the group are not homogeneously distributed amongst individuals (Figures~\ref{fig:Hull}b,~\ref{fig:Voronoi}b, and ~\ref{fig:Voronoi}c). We find that the distribution of internal Voronoi areas is asymmetric about the mode (Figure~\ref{fig:Comp&Exp}) and is robust to the method of measurement (Supplemental Figures 5, 6, and 7). However, the area distribution of non-interacting cells within a two-dimensional random Voronoi networks follows a Gamma distribution with $k=3.6$, which is equivalent to $k_e/k_c \approx 0.6$. As such, deviations from this non-interacting result in Figure~\ref{fig:Comp&Exp}b are a direct result of changing interactions between fish as a result of confinement.

We find that changes in structural and dynamic trends occur between $R=14$cm and $R=34.25$cm such as fish motion (Figure~\ref{fig:msd_Fig}e), modal Voronoi area (Figure~\ref{fig:Voronoi}e), distributions of Voronoi areas (Figure~\ref{fig:Voronoi}f), the shapes of these distributions (Figures~\ref{fig:Comp&Exp}b, \ref{fig:gamma_fit}), and the inferred radii and energy of interaction between fish (Figure~\ref{fig:mc_fit}). As such, we conclude that spatial confinement is a robust method to control the dynamical and mechanical properties of this non-equilibrium material.

\section*{Methods} 
\subsection*{Fish Care}
Cardinal tetra (\textit{Paracheirodon axelrodi}) fish are purchased from the Aquarium Co-Op in Edmonds WA. Fish are housed in a $\sim50$ gallon living aquarium at a maximum density of 1 fish/gallon. Lights are set to a 12h-12h day-night cycle; all experiments are done during the daytime setting. Water is kept at a temperature between $77^\circ$ and $79^\circ$F. Water pH is monitored and adjusted around 7.2-7.4 and ammonia, nitrate, and nitrite levels are monitored using the API Freshwater Test Kit and kept at undetectable levels. Fish are fed once daily. All fish housing and handling is approved by an Institutional Animal Care and Use Committee.

\subsection*{Experimental Details}
For experimental observations, 25 fish are chosen at random from sets of 30-60 fish in all data except Figure~\ref{fig:L_Hist} where 50 fish were used. The fish are transferred to shallow cylindrical arenas made for experiments. The temperature of the observation tank is kept between $77^\circ$ and $79^\circ$F and the water used for this tank is directly taken from the living aquarium. The water depth is kept at 1.5cm with a tolerance of $\pm 0.1$cm across the arena. This minimizes 3D fish crossings that affect our fish identification algorithm. Arenas are made from either custom acrylic or white PVC. Clear acrylic arenas are lined with white tape to match PVC. Arena sizes are radii $R = 8$cm, 11 cm, 14cm, 34.25cm, and 44.25cm.

Shallow arenas are submerged in a large $\sim 200$ gallon water bath with active heating and water circulation which acts as a thermal reservoir but avoids generating any flow in the observation arenas. Water within the shallow container containing the fish is static except when perturbed by the fish within the 
observation arenas. Fish are left undisturbed in the imaging arena for a minimum of one hour before imaging for acclimatization. 

Fish are back-lit by submerged broadband visible light. A diffusive acrylic layer separates the light source and the imaging aquarium base which helps to homogeneously illuminate the field of view. Room lights are turned off during acclimatization and experiments. Videos are recorded from overhead with a Pixelink PL-D7620 machine vision camera at 10 frames per second for up to 60 minutes.

\subsection*{Image analysis}
Videos are taken with lighting optimized to ensure shadows, bubbles, and any other visual noise are minimized before using the open-source software Trex~\cite{walter_trex_2021} to threshold the videos and determine position, velocity, and orientation for individual fish. 
The pixel-to-centimeter conversion is found by taking a photograph of a ruler at the bottom of the arena after each experiment without disturbing the camera setup. 

\subsection*{MSD Analysis}
The mean-squared-displacement MSD is calculated for each fish and averaged for all fish in an experiment. When we lose continuity in fish trajectories due to tracking errors, we ensure that the MSD is only calculated for consecutively tracked frames. The average consecutive track lengths for any given fish range from 59s in the largest $R=44.25$cm arena to 18.8s in the smallest $R=8$cm arena.
Scaling exponent $\alpha$ is calculated via a power-law fit for $\tau \leq 1$s.

\subsection*{Fitting Probability Distributions}
Two parabolas are fit to a probability distribution smoothed with a Gaussian filter and forced through a common peak $A_0$ in data such as Figure~\ref{fig:Comp&Exp}a. Each parabola is either fit to values less than $A_0$ for compression or greater than $A_0$ for expansion. The range of values around $A_0$ that the parabolas are fit to be the same for both and is defined separately for each experiment. This range falls between 1/3 to 2/3 the value of $A_0$.

\section*{Acknowledgements}

APT was supported by the M.J. Murdock Charitable Trust Award Number FSU-201913717. APT, GK, MR, and JS were supported by the National Science Foundation Award Number 2137509. JH, DB, and APT were supported by National Science Foundation Award Number DMR-2046683.

\section*{Author contributions statement}
GK and MR collected experimental data.
GK, MR, and ALR contributed analytical tools.
GK, JH, MR, and JS analyzed data.
JH and DB conducted simulations.
APT, GK, JH, DB, and MR wrote the manuscript.
APT and GK edited the manuscript.
APT designed and conceived the work.

\section*{Additional information}

The authors have no competing interests.

\bibliography{lib0523.bib}
\end{document}